\documentclass[aps,preprint,nofootinbib,preprintnumbers,eqsecnum,superscriptaddress]{revtex4-1}

\usepackage{color}

\newcommand{\nb}[1]{\color{blue}}

\newcommand{\hl}[1]{\color{magenta}}

\usepackage[
	  pagebackref=false,
	  colorlinks=true,
      linkcolor=blue,
      urlcolor=blue,
      filecolor=black,
      citecolor=red,
      pdfstartview=FitV,
      pdftitle={},
        pdfauthor={},
        pdfsubject={},
        pdfkeywords={},
        pdfpagemode=None,
        bookmarksopen=true
      ]{hyperref}

\usepackage[normalem]{ulem}
\usepackage{amsmath}
\usepackage{enumerate}
\usepackage{amsfonts}
\usepackage{epsfig}
\usepackage{mathbbol}

\def\Tr{\mathop{\rm Tr}}

\newcommand\half{{\ensuremath{\frac{1}{2}}}}
\newcommand\p{\ensuremath{\partial}}

\newcommand\vev[1]{{\ensuremath{\left\langle{#1}\right\rangle}}}

\newcommand\ket[1]{\ensuremath{\lvert{#1}\rangle}}
\newcommand\bra[1]{\ensuremath{\langle{#1}\rvert}}

\newcommand{\be}{\begin{equation}}
\newcommand{\ee}{\end{equation}}
\newcommand{\bea}{\begin{eqnarray}}
\newcommand{\eea}{\end{eqnarray}}
\newcommand{\bega}{\begin{gather}}
\newcommand{\eega}{\end{gather}}

\newcommand{\bi}{\begin{itemize}}
\newcommand{\ei}{\end{itemize}}
\newcommand{\ben}{\begin{enumerate}}
\newcommand{\een}{\end{enumerate}}
\newcommand{\bca}{\begin{cases}}
\newcommand{\eca}{\end{cases}}
\newcommand{\bln}{\begin{align}}
\newcommand{\eln}{\end{align}}
\newcommand{\bst}{\begin{split}}
\newcommand{\est}{\end{split}}
\def\ie{\begin{equation}\begin{aligned}}
\def\fe{\end{aligned}\end{equation}}
\newcommand{\bma}{\le(\begin{matrix}}
\newcommand{\ema}{\end{matrix}\ri)}

\newcommand\al{{\alpha}}
\newcommand\ep{\epsilon}
\newcommand\sig{\sigma}

\newcommand\Lam{\Lambda}
\newcommand\om{\omega}

\newcommand\ga{{\ensuremath{{\gamma}}}}
\newcommand\Ga{{\ensuremath{{\Gamma}}}}
\newcommand\de{{\ensuremath{{\delta}}}}

\newcommand\vp{\varphi}
\newcommand\ka{\kappa}

\newcommand\da{{\dagger}}

\def\th{{\theta}}

\newcommand\ov{\over}
\newcommand\ha{{\half}}

\def\le{\left}
\def\ri{\right}

\newcommand\sF{{\ensuremath{{\mathcal F}}}}

\newcommand\sL{{\ensuremath{{\mathcal L}}}}

\newcommand\sP{{\ensuremath{{\mathcal P}}}}

\newcommand\sT{{\mathcal T}}

\newcommand\vx{{\vec x}}
\newcommand\vk{{\vec k}}

\newcommand{\hmu}{{\hat \mu}}

\newcommand{\rmi}{{\rm i}}
\newcommand{\rmj}{{\rm j}}

\newcommand{\local}{{\rm dynamical}}

\begin{document}

\title{The second law of thermodynamics from symmetry and unitarity}

\preprint{MIT-CTP/4859}

\author{Paolo Glorioso}
\affiliation{Kadanoff Center for Theoretical Physics, \\
University of Chicago, Chicago, IL 60637, USA}

\author{Hong Liu}
\affiliation{Center for Theoretical Physics, \\
Massachusetts
Institute of Technology,
Cambridge, MA 02139 }

\begin{abstract}

\noindent  The second law of thermodynamics states that for a thermally isolated system entropy never decreases.
Most physical processes we observe in nature involve variations of macroscopic quantities over spatial and temporal scales much larger than microscopic molecular collision scales and thus can be considered as in local equilibrium. For a many-body system in local equilibrium a stronger version of the second law applies which says that the entropy
production at each spacetime point should be non-negative.  In this paper we provide a proof of the second law for such systems and  a first derivation of the local second law.
For this purpose we develop a general non-equilibrium effective field theory
of slow degrees of freedom from integrating out fast degrees of freedom in a quantum many-body system and consider
its classical limit. The key elements of the proof are the presence of a $Z_2$ symmetry, which can be considered as a proxy for local equilibrium and micro-time-reversibility, and a classical remnant of quantum unitarity.
 The $Z_2$ symmetry leads to a  local current from a procedure analogous to that used in the Noether theorem.
Unitarity leads to a definite sign of  the divergence of the current.
We also discuss the origin of an arrow of time, as well as the coincidence of
causal and thermodynamical arrows of time.
Applied to hydrodynamics, the proof gives a first-principle derivation of the phenomenological entropy current condition and provides a constructive procedure for obtaining the entropy current.

\end{abstract}

\today

\maketitle


\section{Introduction}

Ink dropped into a bowl of water spreads and does not regroup, a broken toy  does not reassemble itself, heat does not pass spontaneously from a cooler to a hotter object; such irreversible phenomena are ubiquitous in nature.
They are explained by the second law of thermodynamics, which associates a physical quantity called
entropy with an equilibrium state of matter and states that {\it for a thermally isolated system entropy never decreases}.
Heuristically speaking, entropy is a measure of manifest disorder.
Ink molecules spreading uniformly in water is more disordered than concentrated in a single drop and thus has a higher
entropy.  The second law governs essentially all aspects of the universe, from molecular dynamics to star formations, from engines to biological systems, from cosmology to black holes and quantum gravity.

For many physical processes in nature, in fact a stronger version of the second law is in operation. Typical physical processes we observe involve variations of macroscopic quantities over spatial and temporal scales much larger than microscopic molecular collision scales. Thus for any region $V$ whose size is much larger than molecular scales but much smaller than the distance and time scales of variations can be considered as  in {\it local equilibrium}. 
 Going to a continuum description we can then introduce an entropy density
$s(t, \vec x)$ and an entropy flow vector $S^i$,  to express the second law in a local form as
\be \label{l2nd}
\p_t \int_V d^3 \vec x \, s (t, \vx) + \int_{\p V} d \vec \sig \cdot \vec S \geq 0, \quad \Rightarrow \quad  \p_\mu S^\mu \geq 0,  \quad S^\mu = (s, S^i)
\ee
 We emphasize that local equilibrium does not mean near equilibrium as variations of macroscopical physical quantities can be big over large distances and long time periods, and in fact includes most far-from-equilibrium situations observed in nature. 

While the second law was first formulated by Clausius more than one and a half centuries ago,
understanding how it arises from basic laws of physics which are time symmetric, remains incomplete. 
In particular, while the local second law has played a central role in formulating many phenomenological theories including fluid mechanics~\cite{LL}, there has not been any derivation of it from first principle.

In this paper we provide a proof of the second law and  the local second law~\eqref{l2nd} 
for the classical limit of any quantum many-body system in local equilibrium. The second law is proved in general while the local version is proved perturbatively in a derivative expansion. The basic idea is as follows. We start by formulating a general non-equilibrium effective field theory for a quantum many-body system obtained by integrating out fast degrees of freedom. Interestingly, unitarity of quantum time evolution imposes various constraints on the action which survive in the classical limit. For example, the action is in general complex and the imaginary part of the action is non-negative. We then impose a $Z_2$ symmetry~\cite{janssen1,janssen2,Sieberer2,CGL,CGL1},   which can be considered as a proxy for local equilibrium and micro-time-reversibility. It implies that the first law
of thermodynamics, Onsager relations, as well as fluctuation-dissipation relations are satisfied locally.
From a procedure analogous to that in the proof of Noether's theorem, the $Z_2$ symmetry leads to a local current which is not conserved,  but whose divergence can be expressed in terms of the imaginary part of the action, and is thus non-negative. At zeroth order in derivative expansion the current is conserved and recovers the standard thermodynamic entropy.
We also discuss the origin of the arrow of time which can be attributed to whether the local equilibrium is established in the past or in the future.

Our proof of the second law complements the existing proofs and brings a number of immediate conceptual implications. 
The celebrated Boltzmann's H theorem~\cite{boltz} applies to dilute gases, and the fluctuation theorems~\cite{boch,jarz1,crooks} apply to classical Hamiltonian systems initially in thermal equilibrium perturbed by external mechanical forces.
Ours applies to all systems in local equilibrium including liquids, critical systems, and quantum liquids such as superfluids and strongly correlated systems. 
While the H-theorem starts with a statistical definition of entropy, here the concept of thermodynamical entropy is emergent, arising 
from a $Z_2$ symmetry. The derivation only shows that there is a monotonic quantity, which turns out to coincide with our usual notion of thermodynamic entropy.
Our derivation also highlights the importance of
quantum unitarity in the monotonicity of entropy evolution. It implies  that if one just writes down a most general classical effective action for {\it a dissipative open system},  entropy evolution may not be monotonic. 

As with earlier derivations~\cite{boltz,boch,jarz1,crooks},  here a thermodynamic arrow of time arises with a choice of boundary condition in time. Without such an extra input the most one could get from an underlying time-symmetric system is the monotonicity of entropy evolution. Also as in other derivations microscopic time reversibility plays a key role.

 Applied to the hydrodynamical action recently proposed in~\cite{CGL,CGL1}, our derivation of the local second law~\eqref{l2nd}  gives a first-principle derivation of the phenomenological entropy current condition and a constructive procedure for obtaining the entropy current.
Recent advances in understanding the entropy current condition in hydrodynamics include~\cite{Banerjee:2012iz,Jensen:2012jh,Bhattacharyya:2013lha,Bhattacharyya:2014bha}. At ideal fluid level, entropy current
arises as a topologically conserved current in the formulation of~\cite{Dubovsky:2011sj,Dubovsky:2005xd}, while in~\cite{deBoer:2015ija,CGL} it arises as the Noether current for an accidental continuous  symmetry. In~\cite{Haehl:2015pja} entropy current is proposed at non-dissipative level as the Noether current associated to a $U(1)$ symmetry, which 
was further advocated in~\cite{Haehl:2015uoc} as a symmetry for full dissipative fluids.

The plan of the paper is as follows. In next section we formulate a general class of non-equilibrium effective field theories. We discuss constraints from unitarity and introduce a $Z_2$ symmetry to impose local equilibrium. In Sec.~\ref{sec:theo} we present a proof of the second law and a perturbative proof of its local second version. We conclude in Sec.~\ref{sec:arrow} with a discussion of the origin of an arrow of time.
We have included a number of appendices which contain explicit examples as well as various background materials and technical details.

\section{Non-equilibrium effective theories}  \label{sec:NEQFT}

In this section we formulate a most general non-equilibrium effective theory obtained from consistently integrating out ``fast'' degrees of freedom in a state of local equilibrium. By fast degrees of freedom we mean either gapped modes or modes 
with a finite lifetime in the long wavelength limit (which can include gapless modes). 
This means that there is a separation of scales between the integrated-out fast and remaining ``slow'' degrees of freedom,
and thus the effective action for the slow modes must be local, i.e. has a regular local expansion in terms of the number of derivatives. The expansion in derivatives is controlled by a small parameter $\ep/L \ll 1$ where $L$ is the characteristic wavelength   (or inverse frequency) of the slow degrees of freedom while $\ep$ is some microscopic scale characterizing the life-times and correlation lengths of fast degrees of freedom.

Consider the path integral for describing expectation values  in a quantum state, which can be defined
on a closed time path  (CTP) contour~\cite{schwinger,keldysh,Feynman:1963fq},  
\bln \label{gen0}
e^{W [\phi_1, \phi_2]}  & = \Tr \le( U (+\infty, -\infty; \phi_1) \rho_0 U^\da (+\infty, -\infty; \phi_2) \ri)
 =   \int_{\rho_0} D \psi_1 D \psi_2 \, e^{i S_0 [\psi_1, \phi_1] - i S_0 [\psi_2; \phi_2]}
\end{align}
where $\rho_0$ denotes the initial state of the system, and $U (t_2, t_1; \phi)$ is the evolution operator of the system
from $t_1$ to $t_2$ in the presence of external sources collectively denoted by $\phi$.   The sources are taken to be slowly varying functions and there are two copies of them, one for each leg of the CTP contour. The second equality is the ``microscopic'' path integral description, with $\psi_{1,2}$  denoting microscopic dynamical variables for the two copies of spacetime of the CTP and $S_0 [\psi; \phi]$ the microscopic action. Now suppose in this system there is a natural separation of slow and fast degrees of freedom and integrate out fast variables, after which
\be
e^{W [\phi_1, \phi_2]}  =  \int D \chi_1 D \chi_2 \, e^{i I_{\rm eff} [\chi_1, \phi_1; \chi_2 , \phi_2; \rho_0]}   \
\label{left}
\ee
where $\chi_{1,2}$ denote slow variables and there are again two copies of them.
It is convenient to introduce the so-called $r-a$ variables~\cite{Chou:1984es} 
\be \label{rava}
\chi_r = \ha (\chi_1 + \chi_2) , \quad \chi_a = \chi_1 - \chi_2, \quad  \phi_r = \ha (\phi_1 + \phi_2) , \quad \phi_a = \phi_1 - \phi_2\
\ee
where as usual $\chi_r$ correspond to physical observables while $\chi_a$ correspond to noises.

Unitarity of time evolution in~\eqref{gen0} imposes nontrivial constraints on $I_{\rm eff}$.
Taking the complex conjugate of~\eqref{gen0} we find that  $W$ satisfies $W^* [\phi_1, \phi_2] = W[\phi_2,  \phi_1]$ which in turn requires that
\be\label{fer1}
  I^*_{\rm eff} [\chi_r , \phi_r; \chi_a, \phi_a] = - I_{\rm eff} [\chi_r , \phi_r; -\chi_a, -\phi_a]   \
\ee
where for definiteness we have taken $\chi_{1,2}$ and sources $\phi_{1,2}$ to be real.
Equation~\eqref{fer1} implies that terms in $ I_{\rm eff}$ which are {\it even} in $a$-variables 
must be pure imaginary. Note that the original factorized form of the action in~\eqref{gen0} is real and is odd in $a$-variables.
 Given that one expects even terms will generically be generated when integrating out fast variables, $ I_{\rm eff}$ is thus generically complex.
Now that $i I_{\rm eff}$ can be real, there is a danger that the integrand of~\eqref{left} can be exponentially increasing with
$\chi_{1,2}$ thus making the path integrals ill-defined. Note, however, that  unitarity of evolution operator $U$ implies\footnote{We thank J.~Maldacena for a comment which led to our better understanding on this.}  
\be \label{pos}
{\rm Im} \, I_{\rm eff} \geq 0 \
\ee
for any dynamical variables $ \chi_{1,2}$ and sources $\phi_{1,2}$.
There is one further constraint from unitarity: in~\eqref{gen0} taking $\phi_1 = \phi_2 = \phi$ and $\chi_1 = \chi_2 = \chi$, we
should have 
\be \label{key1}
I_{\rm eff} [\chi, \phi; \chi, \phi] = 0, \quad {\rm or} \quad
I_{\rm eff} [\chi_r =\chi, \phi_r = \phi; \chi_a=0, \phi_a =0] = 0 \ .
\ee
Equation~\eqref{key1} implies that any term in the action must contain at least one factor of $a$-type variables ($\phi_a$ or $\chi_a$). We give a derivation of~\eqref{fer1}--\eqref{key1} in Appendix~\ref{app:a}.

There are three different regimes for~\eqref{left}. The first is the full quantum level where path integrations describe both quantum and classical statistical fluctuations. The second is the classical limit with $\hbar \to 0$ where the path integrals remain
and describe classical statistical fluctuations. The third is the level of equations of motion from $I_{\rm eff}$ which corresponds to
the thermodynamic limit with all classical and quantum fluctuations neglected.
The consequences~\eqref{fer1}--\eqref{key1} of quantum unitarity concern with general structure of $I_{\rm eff}$ and thus survive in all regimes. In this paper we will work at the level of equations of motion.

It is straightforward to write down the most general effective action $I_{\rm eff}$ for slow modes corresponding to non-conserved quantities.\footnote{Examples of such slow modes include order parameters near a phase transition or gapless modes near a Fermi surface.} For hydrodynamic modes associated with conserved quantities such as the energy-momentum tensor or
 charges of some internal symmetries,  the problem is much trickier both in terms of identifying the appropriate dynamical variables $\chi$ and the symmetries that $I_{\rm eff}$ should obey, and
 has only been recently solved in~\cite{CGL,CGL1} (see also~\cite{Dubovsky:2005xd,Dubovsky:2011sj,Grozdanov:2013dba,Kovtun:2014hpa,Harder:2015nxa,Haehl:2015uoc} for other recent discussions). Here we follow the formulation of the classical limit in~\cite{CGL1}. For definiteness, let us consider a system with a $U(1)$ symmetry.
The most general Lagrangian density for $I_{\text{eff}} = \int
d^d x \, \mathcal L_{\text{eff}}$,  including both non-conserved and hydrodynamic modes can be written as
\be \label{w3}
\sL_{\rm eff} = \sum_{n=1}^\infty  i^{\eta_n}  f^{(n)} [\Lam_r]  \Phi_a^{n} , \qquad \eta_n = \bca 1 & n\; {\rm even} \cr
             0 & n \; {\rm odd}
             \eca \
\ee
where $\Lam_r$ denotes the collection of $r$-variables while $\Phi_a$ denotes the collection of $a$-variables.
The $n$-th term in~\eqref{w3} should be understood as
\be \label{uips}
f^{(n)} [\Lam_r] \Phi_a^n  =  f^{(n)}_{\al_1 \cdots \al_n}
(\Lam_r; \p_\mu) \, \Phi_{a \al_1} (x)  \cdots \Phi_{a \al_n} (x)\
\ee
where  $ f^{(n)}_{\al_1 \cdots \al_n} (\Lam_r; \p_\mu)$  is a function of $\Lam_r$, their derivatives, as well as derivative operators acting on $\Phi_{a \al}$, with index $\al$ running over different $a$-variables.
In~\eqref{w3} the sum starts at $n=1$ due to~\eqref{key1} and the even terms are pure imaginary due to~\eqref{fer1}.
While in practice one only needs to keep the first few terms in~\eqref{w3} in powers of $a$-variables and derivatives,
here we keep the full dependence for both.
More explicitly, we can write $\Lam_r = \{\chi_{r \rmi}, \beta^\mu , \hmu\}$ and $\Phi_a = \{\chi_{a \rmi}, \p_\mu X_{a \nu}, \p_\mu \vp_a \}$, where $\chi_{r \rmi}, \chi_{a \rmi}$ denote
slow variables for non-conserved quantities with index $\rmi$ labeling different species.  $\beta^\mu = \beta (x) u^\mu (x)$ and $\hmu = \beta (x) \mu (x)$ are  hydrodynamical $r$-variables  where $\beta (x), u^\mu, \mu (x)$ are respectively local inverse temperature, local velocity field and local chemical potential. Hydrodynamical $a$-variables  $X_a^\mu, \vp_a$  correspond to noises for the energy-momentum and the $U(1)$ charge, and must always be accompanied by derivatives as already indicated in $\Phi_a$. In particular, in derivative expansion, $ \p_\mu X_{a \nu}, \p_\mu \vp_a$ should be counted as having zero derivatives.
For definiteness we use the relativistic regime throughout the paper. For a non-relativistic system the discussion is completely parallel.

Let us now turn to equations of motion of $I_{\rm eff}$. Given that~\eqref{w3} contains at least one factor of $\Phi_a$, equations of motion from varying with respect to any $r$-variables can be consistently solved by setting all $a$-variables to zero. Thus nontrivial equations of motion come from varying with respect to $a$-variables and furthermore only $f^{(1)}$ are relevant. More explicitly let us write
\be \label{eom11}
 f^{(1)} [\Lam_r] \Phi_a = E_{\rmi} \chi_{a \rmi} + T^{\mu \nu} \p_\mu X_{a \nu} + J^\mu \p_\mu \vp_a
 \ee
 where by using integration by parts we can move all derivatives on $\Phi_a = \{\chi_{a \rmi}, \p_\mu X_{a \nu}, \p_\mu \vp_a \}$
 to the other factors. Thus the equations of motion can be written as
 \be \label{eom12}
 E_{\rmi} =0, \qquad \p_\mu T^{\mu \nu} =0, \qquad \p_\mu J^\mu =0  \ .
 \ee
$T^{\mu \nu}$ and $J^\mu$ can be interpreted as the (macroscopic) energy momentum tensor and $U(1)$ current, and the second and third equations are simply the corresponding conservation equations. In Appendix~\ref{app:a7} we discuss two explicit examples, model A for critical dynamics and fluctuating hydrodynamics for a relativistic charged fluid.

We now impose a $Z_2$ symmetry, to which we will refer as the dynamical KMS symmetry. More explicitly we require that
\be \label{lkms}
I_{\rm eff} [\Lam_r, \Phi_a] =I_{\rm eff}   [\tilde \Lam_r, \tilde \Phi_a]
\ee
where in the classical limit $\hbar \to 0$
\be  \label{glkms}
\tilde \Lam_{r} (-x) = \Lam_{r} (x),  \qquad  \tilde \Phi_{a} (-x) = \Phi_{a} (x) + i \Phi_{r} (x)
\ee
with $\Phi_{r} = \{\beta^\mu \p_\mu \chi_{r \rmi},   \p_\mu \beta_\nu  , \p_\mu \hmu\}$ respectively for
$\Phi_a = \{\chi_{a \rmi}, \p_\mu X_{a \nu}, \p_\mu \vp_a \}$.\footnote{In writing down~\eqref{glkms} for definiteness we have taken that  microscopically the system is invariant under $\sP \sT$ (not necessarily separate $\sT$ or $\sP$)
and  the phases for $\chi$ under $\sP \sT$ to be $1$. One can easily adapt~\eqref{glkms} to write down the corresponding transformations for a system with only microscopic $\sT$ invariance. \label{foot:2}} The dynamical KMS symmetry plays the role of imposing
micro-time-reversibility and local equilibrium. In Appendix~\ref{app:a2} we elaborate more on their motivations and in Appendix~\ref{app:a7} we illustrate their physical implications using some examples. For non-conserved quantities $\chi_{r \rmi}, \chi_{a \rmi}$ the transformation~\eqref{glkms} generalizes to local equilibrium a previously known transformation
 characterizing a thermal ensemble~\cite{janssen1,janssen2,Sieberer2}. Those for hydrodynamical variables have only been recently proposed in~\cite{CGL,CGL1}. 
The transformations~\eqref{glkms}, to which we will refer as dynamical KMS transformations below,  are $Z_2$. This is obvious for $\Lam_r$. For $\Phi_a$, we have
\be
\tilde {\tilde \Phi}_a = \tilde \Phi_a (-x) + i \tilde \Phi_r (-x) = \Phi_a (x)  + i  \Phi_r (x) - i \Phi_r (x)  = \Phi_a (x)
\ee
where we have used that $\tilde \Phi_r (-x) = - \Phi_r (x)$ due to the single derivative inside $\Phi_r$.

Since $\Phi_{r }$ contains one derivative, the dynamical KMS condition~\eqref{lkms} relate $n$-th derivative terms in $f^{(1)}$ with $(n-1)$-th derivative terms in $f^{(2)}$, $(n-2)$-th derivative terms in $f^{(3)}$, etc., all the way to zeroth derivative terms in $f^{(n)}$.
Thus even though the equations of motion~\eqref{eom12} only involve $f^{(1)}$, all terms in~\eqref{w3} play a role through~\eqref{lkms}.  Also note that the zero derivative part of $f^{(1)}$ should be invariant by itself, which we will see have interesting implications.

One can include external sources in~\eqref{w3}. For $\phi_1 =\phi_2 = \phi$, i.e. $\phi_r = \phi, \phi_a =0$ one can simply replace $f^{(n)}$ in~\eqref{w3} by $f^{(n)} [\Lam_r, \phi]$, i.e. now becoming also local functions of $\phi$ and their derivatives.
Turning on $\phi_r$  for the stress tensor corresponds to putting the system on a curved spacetime metric $g_{\mu \nu}$ in which case one should replace all derivatives by covariant derivatives associated with $g_{\mu \nu}$.
Also note that the dynamical KMS transformation for $\phi$ is $ \tilde \phi_{r} (-x) = \phi_{r} (x)$. We will not need to turn on $\phi_a$ for this paper.


\section{A theorem of local second law of thermodynamics} \label{sec:theo}

In this section we will prove a general theorem on the second law of thermodynamics.

{\it Theorem: }  for any local effective theory~\eqref{w3}  which satisfies~\eqref{pos}
and the $Z_2$ dynamical KMS symmetry~\eqref{lkms}
there exists a local density $S^0$ which satisfies (given equations of motion)
\be 
 \Delta S  \equiv  \int_{t=t_2} d^{d-1}x \, S^0-\int_{t=t_1} d^{d-1}x \, S^0   \geq 0\ .
\ee
Furthermore perturbatively to all orders in derivative expansion there exists a local current $S^\mu$ satisfying
\be \label{th1}
\p_\mu S^\mu \geq 0  \ .
\ee
Denote $S^\mu_0$ as the expression for $S^\mu$ at zeroth order in derivative expansion, then
\be \label{th2}
\p_\mu S^\mu_0 = 0 \
\ee
and recovers the standard equilibrium thermodynamical entropy.
The theorem can be straightforwardly generalized to include diagonal external sources $\phi$. In particular, in a curved spacetime metric $g_{\mu \nu}$  one should replace the derivatives in~\eqref{th1} and~\eqref{th2} by covariant derivatives associated with $g_{\mu \nu}$. Below for notational simplicity we will present the proof without external sources.

\subsection{Proof of the second law}

We start by  deriving some general consequences of the dynamical KMS symmetry~\eqref{lkms}, which implies that
\be \label{ryt}
\sL _{\rm eff}= \tilde \sL_{\rm eff} - \p_\mu V^\mu
\ee
where $\tilde \sL_{\rm eff}$ is obtained by plugging~\eqref{glkms} into~\eqref{w3} and taking $x \to - x$, i.e.
\be
\tilde \sL_{\rm eff} = \sum_{n=1}^\infty  i^{\eta_n} f^{(n) *} [\Lam_r] (\Phi_a + i \Phi_r)^n    \
\ee
with $f^{(n) *}_{\al_1 \cdots \al_n} (\Lam_{r} (x); \p_\mu) \equiv  f^{(n)}_{\al_1 \cdots \al_n} (\Lam_{r} (x); - \p_\mu)  $.
Expanding $V^\mu$ in terms of the number of $a$-fields
\be
V^\mu = \sum_{n=0}^\infty 
 i^{\eta_n} V^\mu_n 
\ee
with $V^\mu_n$ containing $n$ factors of $\Phi_a$,  we find ($[x]$ denotes the integer part of $x$)
\bega \label{gkm1}
\sum_{n=1}^\infty 
 (-1)^{[{n\ov 2}]} f^{(n)*} [\Lam_r] \Phi_r^n
 = \p_\mu V^\mu_0
  \end{gather}
and for $k \geq 1$ 
\be
 \label{gkm2}
 f^{(k)} \Phi_a^k  + \p_\mu V^\mu_k
 =  f^{(k) *} \Phi_a^k
+
 \sum_{n=1}^\infty \ep_{nk} 
 \sum_{j_1, \cdots j_n}
 f^{(n+k) * }_{\al_1 \cdots \al_{j_1} \cdots \al_{j_n} \cdots \al_{k+n}}
 \Phi_{a \al_1} \cdots \Phi_{r \al_{j_1}} \cdots \Phi_{r \al_{j_n}} \cdots \Phi_{a \al_{k+n}}
\ee
where $ \Phi_{a \al_1} \cdots \Phi_{r \al_{j_1}} \cdots \Phi_{r \al_{j_n}} \cdots \Phi_{a \al_{k+n}}$ denotes that at
$1 \leq j_1 < j_2 < \cdots j_n \leq n+k$, $\Phi_{a\al}$ are replaced by the corresponding $\Phi_{r\al}$.
We then sum over all possible replacements. In~\eqref{gkm2} $\ep_{nk}$ is given by
\be
\ep_{nk} = \bca (-1)^{[{n \ov 2}]} & k \; {\rm even}
 \cr (-1)^{[{n+1 \ov 2}]} & k \;  {\rm odd} \eca \ .
\ee
Since~\eqref{gkm2} should apply for any $\Phi_a$, by replacing $\Phi_{a\al}$ in~\eqref{gkm2} by
the corresponding $\Phi_{r \al}$ we  find that
\be \label{meq}
 f^{(k)} \Phi_r^k  + \p_\mu \hat V^\mu_k = f^{(k)*}  \Phi_r^k   + \sum_{n=1}^\infty \ep_{nk} C_{n+k}^n f^{(n+k)*}  \Phi_r^{n+k}
 \ee
 where $ \hat V^\mu_k$ is obtained from $V^\mu_k$ by replacing  all $\Phi_{a}$ by
the corresponding $\Phi_{r}$.
It can be shown that in~\eqref{gkm2} $V_k^\mu$ for $k \geq 1$ can be set to zero by absorbing total derivative terms into the definition of $\sL_{\rm eff}$ (see Appendix~\ref{app:a3}). This is not possible for $V_0^\mu$ in~\eqref{gkm1} as $\sL_{\rm eff}$ does not contain any term with no $\Phi_a$ factors. For $k=1$ we will need to perform a further integration by parts to write the Lagrangian in the form of~\eqref{eom11}, which can generate a nonzero $V_1^\mu$.  Below for notational simplicity we assume such a ``canonical'' Lagrangian has been chosen~(i.e. with only possible nonzero $V_0^\mu$ and $V_1^\mu$).

Using equation~\eqref{meq} to solve for $ f^{(k)*}  \Phi_r^k$ and substituting the resulting expressions into~\eqref{gkm1} we find that
 \be
 \p_\mu W^\mu 
=  \sum_{k=1}^\infty \ep_k  f^{(k)} \Phi_r^k
 \label{ghr}
 \ee
 where
 \be
W^\mu = V^\mu_0 - \hat V_1^\mu  , \qquad  \ep_k =  (-1)^{[{k-1 \ov 2}]} \ .
 \ee
From~\eqref{eom11}, the $k=1$ term in~\eqref{ghr} has the form
\be \label{eom13}
 f^{(1)} [\Lam_r] \Phi_r = E_{\rmi} \p_0 \chi_{r\rmi}+ T^{\mu \nu} \p_\mu \beta_{\nu} + J^\mu \p_\mu \hmu
 = \p_\mu \le(T^{\mu \nu} \beta_{\nu}  + \hmu J^\mu \ri)
 \ee
where in the second equality we have used the equations of motion~\eqref{eom12}.
We thus  can write~\eqref{ghr} as
 \be \label{cse}
 \p_\mu S^\mu = \sum_{k=2}^\infty \ep_k  f^{(k)} \Phi_r^k, \qquad S^\mu = V^\mu_0 - \hat V_1^\mu - T^{\mu \nu} \beta_\nu - \hmu J^\mu  \ .
 \ee
Since the right hand side of~\eqref{cse} starts at second order in derivatives we immediately conclude that at zeroth order in derivatives
\be \label{s0}
\p_\mu S_0^\mu = 0 \ .
\ee
Equation~\eqref{s0} can also be understood as follows. At zeroth derivative order the $f^{(1)}$ term should be invariant under~\eqref{glkms} by itself, and given that it is linear in $\Phi_a$, this implies that the $f^{(1)}$ term is invariant under a continuous ``accidental''  symmetry
\be
\Phi_a (x) \to \Phi_a (x) + i \ep \Phi_r (x)
\ee
where $\ep$ is an arbitrary infinitesimal constant.  $S_0^\mu$ can be then identified as the Noether current for this symmetry. For an ideal fluid this was observed earlier in~\cite{CGL}. This continuous symmetry has also been proposed to describe general non-dissipative~\cite{Haehl:2015pja} and dissipative fluids~\cite{Haehl:2015uoc}. 

 We will now show that the quantity on the right hand side of~\eqref{cse} is non-negative. Expanding both sides of~\eqref{cse} in derivatives, at $n$-th order we have
 \be \label{cse1}
 \p_\mu S_{n-1}^\mu = \sum_{k=2}^{n} \ep_k f^{(k, n-k)} \Phi_r^k  = \sum_{k=2}^{n} \ep_k F^{(k,n-k)}, \quad F^{(m,k)} \equiv f^{(m,k)} \Phi_r^m
 \ee
 where the second upper index of $f$ denotes the number of derivatives.  Note that $F^{(m,k)}$ has $m+k$ derivatives.
 Since from~\eqref{pos} only $f^{(k)}$ with $k$ even has non-negative properties, in~\eqref{cse1} we would like now to express $f^{(k)}$ with  odd $k \geq 3$  in terms of those with even $k$'s.
This can be achieved by examining~\eqref{meq} with an odd $k=2m+1$ at each derivative order. More explicitly for $m \geq 1, l \geq 0$ we have\footnote{Equation~\eqref{c2} with $m=1$ is obtained from~\eqref{meq} with $k=1$.
There is no $\p_\mu \tilde V_1^\mu$ as one can show that $\p_\mu V_1^\mu$ can only have odd number of derivatives. See Appendix~\ref{app:a3}.}
\bega \label{c1}
F^{(2m+1,2l+1)} + \ha \sum_{n=1}^l (-1)^n C_{2m+1+2n}^{2n} F^{(2m+1+2n,2l+1-2n)}
=\ha  \sum_{n=0}^l (-1)^{n+1} C_{2m+2n+2}^{2n+1} F^{(2m+2n+2,2l-2n)} \\
\sum_{n=0}^l (-1)^n C_{2m+1+2n}^{2m-1} F^{(2m+1+2n,2l-2n)} = \sum_{n=0}^{l} (-1)^{n} C_{2m+2n}^{2m-1} F^{(2m+2n,2l-2n+1)}
\label{c2}
\end{gather}
where we have used $f^{(m,2l+1)} = - f^{(m,2l+1)*}$ and $f^{(m,2l)} = f^{(m,2l)*}$.
One can also obtain two other sets of equations using~\eqref{meq} with even $k$. It can be checked they are equivalent to~\eqref{c1}--\eqref{c2}.

Equations~\eqref{c1}--\eqref{c2} are two sets of linear equations which can be used to solve for $F^{(2m+1, 2n+1)}$ and $F^{(2m+1, 2n)}$ with $m \geq 1$ in terms of $F^{(2m, k)}$. Here we give the final answer leaving the details to Appendix~\ref{app:a4},
\bega \label{Ff1}
F^{(2m+1,2n+1)}  =  {1 \ov 2m+1} \sum_{k=0}^{n}  (-1)^{k}   C_{2m+2k+2}^{2m} G_{2 k+2}  F^{(2m+2k+2, 2n-2k)}  , \\
\label{Ff2}
F^{(2m+1, 2n)} = 
 {2 \ov 2m+1} \sum_{k=0}^{n} ( -1 )^{ k }  C_{2m + 2k}^{2k} B_{2k} F^{(2m+2k, 2n-2k+1)}  \ .
\end{gather}
where $B_n$ and $G_n$ are Bernoulli and Genocchi numbers respectively. Plugging~\eqref{Ff1}--\eqref{Ff2} into~\eqref{cse1} we find for $n \geq 1$ (see Appendix~\ref{app:a4} for details)
\bega \label{c4}
  \p_\mu  S_{2n-1}^\mu  =
   \sum_{k=1}^n (-1)^k G_{2k}  F^{(2k, 2n-2k)}  , \qquad
 \p_\mu S_{2n}^\mu = \sum_{k=1}^n 2 (-1)^{k+1} B_{2k}  F^{(2k, 2n-2k+1)}
\end{gather}
where the subscript in $S^\mu$ denotes the number of derivatives.

Combining the above equations 
we find that
\be \label{fen}
\p_\mu S^\mu = \sum_{k=1}^\infty (-1)^k G_{2k} F^{(2k,{\rm even})} + \sum_{k=1}^\infty 2 (-1)^{k+1} B_{2k}
F^{(2k, {\rm odd})}
\ee
where $F^{(2k,{\rm even})} = \sum_{n=k}^\infty F^{(2k,2n-2k)}  $ denotes the sum of all even derivative terms in $F^{(2k)}$ and $F^{(2k,{\rm odd})} = \sum_{n=k}^\infty
F^{(2k, 2n-2k+1)}  $ is the sum of odd derivative terms.
Using the integral representations of $B_{2k}, G_{2k}$, equation~\eqref{fen} can be written as (see Appendix~\ref{app:a6})
\be
\begin{split} \label{yuo}
\p_\mu S^\mu &= \pi \int_{-\infty}^\infty \frac{1}{\sinh^2 (\pi z)}\left(\cosh(\pi z)f_E(z)+f_O(z)\right)dz
\end{split}\ee
with
\be f_E(z) \equiv \sum_{k=1}^\infty F^{(2k,\text{even})}z^{2k},\quad f_O(z) \equiv \sum_{k=1}^\infty F^{(2k,\text{odd})}z^{2k}\ .
\ee

Now introduce 
\be 
F_E(z)=\int_{t_1}^{t_2}dt\int d^{d-1}x \, f_E(z), \qquad F_O(z)=\int_{t_1}^{t_2}dt \int d^{d-1}x \, f_O(z) \ .
\ee
Equation~\eqref{pos} and its more refined version (\ref{posit1}) imply that  $F_E(z) + F_O(z) \geq 0$. Since this holds for any $\Phi_r$, the same inequality holds with sign of derivatives flipped, $F_E(z)-F_O(z)\geq 0$. We thus conclude
\be\label{fefo}
F_E(z)\geq |F_O(z)|\geq 0 \ .
\ee

Integrating (\ref{yuo}) over spacetime then leads to
\bea 
 \Delta S & \equiv & \int_{t=t_2} d^{d-1}x \, S^0-\int_{t=t_1} d^{d-1}x \, S^0  \cr
& = & \pi \int_{-\infty}^\infty \frac{1}{\sinh^2 (\pi z)}\left((\cosh(\pi z)-1)F_E(z)+F_E(z)+F_O(z)\right)dz
\geq 0\ .
\label{yuo2}
\eea
This proves the second law for a system in local equilibrium.  Note that (\ref{yuo2}) does not rely on the derivative expansion nor $\Phi_a$ expansion perturbatively and should be applicable if one is able to resum these series.\\

\subsection{Proof of the local second law}

We now show that the local second law holds  perturbatively in derivative expansion. 

First from the discussion of Appendix~\ref{app:lag}  equation~\eqref{pos} implies that 
\be \label{imo}
{\rm Im} \le(\sL_{\rm eff }\ri)_0 \geq 0
\ee
where $ \le(\sL_{\rm eff }\ri)_0$ denotes the zero derivative part of $\sL_{\rm eff }$. From~\eqref{w3} we then find that 
\be \label{rru}
 \sum_{n=1}^\infty f^{(2n, 0)} \Phi_a^{2n}  \geq 0 
 \ee
where again the second upper  index of $f$ denotes the number of derivatives. Now replace $\Phi_a$  by $\Phi_r$. As $\Phi_r$ contains one derivative, equation~\eqref{rru} then implies that 
\be \label{posit2} 
f^{(2,0)}\Phi_r^2\geq 0\  .
\ee
We will assume below that the quadratic form $f^{(2,0)}$ is invertible.

We can now write~\eqref{fen} (or~\eqref{cse}) as 
\be \label{hum}
\p_\mu S^\mu = f^{(2,0)}\Phi_r^2  +H_{(3)}+H_{(4)}+\cdots \ ,
\ee
where $H_{(n)}$ denotes a scalar with $n$ derivatives. Note that all $H_{(n)}$ contain at least two factors of $\Phi_r$ (with possible  derivatives acting on them). By integrating by parts we can always isolate a single factor of $\Phi_r$ with no derivative acting on it, i.e. we can write 
\be\label{Hn} 
H_{(n)}=2\Phi_r h_{(n-1)}+\p_\mu K_{(n-1)}^\mu,
\ee
where $h_{(n-1)}$ is a tensor of derivative order $n-1$, and $K_{(n-1)}^\mu$ is a vector of derivative order $n-1$. 
Now the total derivative derivative term of~\eqref{Hn} can be absorbed into the left hand side by redefining $S^\mu$, while the first term of~\eqref{Hn} can always be combined with $ f^{(2,0)}\Phi_r^2$ into a square order by order in derivative expansion~\cite{Bhattacharyya:2013lha}. More explicitly 
\be 
 f^{(2,0)}\Phi_r^2  + 2\Phi_r h_{(2)} =  f^{(2,0)} (\Phi_r +  \tilde h_{(2)} )^2 - f^{(2,0)}  \tilde h_{(2)}^2 , \quad
 \tilde h_{(2)} = \le(f^{(2,0)}\ri)^{-1} h_{(2)}  \ 
  \ee
and the leftover term $f^{(2,0)}  \tilde h_{(2)}^2 $ again contains at least two factors of $\Phi_r$ (with possible  derivatives acting on them) and can be absorbed to $H_4$. Carrying this procedure to $H_n$, the right hand side of~\eqref{hum} then becomes 
\be 
 f^{(2,0)} (\Phi_r +  \tilde h_{(2)} + \cdots +  \tilde h_{(n-1)} )^2 + H_{n+1} + \cdots  \ .
\ee
We thus have proved that order by order in derivative expansion 
\be \label{len}
\p_\mu S^\mu \geq 0 \ . 
\ee
Note for this perturbative proof of~\eqref{len} the detailed structure of~\eqref{fen} is not needed. It does not matter whether one starts from~\eqref{fen} or~\eqref{cse}.

The above discussion provides a constructive procedure to obtain the explicit expression of the entropy current order by order in derivative expansion. 

If instead of~\eqref{imo} we have 
\be \label{poo}
{\rm Im}\, \sL_{\rm eff}  \geq 0
\ee
then using argument leading to~\eqref{fefo} we have
\be\label{fefo1}
f_E (z)\geq |f_O(z)|\geq 0 \
\ee
which then leads to~\eqref{len} {\it non-perturbatively.} 
As discussed in Appendix~\ref{app:lag}, however, one in general cannot conclude~\eqref{poo} from~\eqref{pos}. This still leaves the possibility for a given ${\rm Im} \, I_{\rm eff}$ whether one could always choose a particular ${\rm Im} \, \sL_{\rm eff}$ by using the freedom of adding total derivatives such that~\eqref{poo} holds. It is not clear to us whether this is possible. Even if this is possible, it is not clear to us what the precise physical implication is. Note that neither our proof of~\eqref{yuo2} nor the perturbative proof of~\eqref{len} depends on choice of such total derivatives.

As in the usual Noether procedure, the choice of $S^\mu$ is not unique since it can be modified by adding total derivatives to the Lagrangian. We stress that the equilibrium part of $S^\mu$, i.e. the part with zero derivatives, is unique, as this part of the Lagrangian is not affected by adding total derivatives.

Applying the explicit expression~\eqref{cse} for $S^\mu$ to explicit examples, at zeroth derivative orders, which means one can
ignore spacetime variations, we find $S^\mu_0$ recovers the standard thermodynamic entropy density. In particular,  applying it to
hydrodynamics, we find the usual ideal fluid form $S_0^\mu = \beta p_0  -  T^{\mu \nu} \beta_\mu -  J^\mu \hmu $ where $p_0$ is the pressure density.\footnote{Note that the proof of~\eqref{th1} does not depend on the specific form of~\eqref{glkms} nor the $Z_2$ nature of it.  But the identification of $S^\mu$ with entropy current does depend on it.} See Appendix~\ref{app:a7} for more details.

\section{Discussions} \label{sec:arrow}

We have presented a derivation of the local second law of thermodynamics, which implies an arrow of time.
The process of integrating out fast degrees of freedom, while generally generating dissipative terms,  does not introduce an arrow of time, as the signs of the dissipative terms can be either way. So the arrow of time must be generated from the only other input of our proof, the $Z_2$ symmetry. Indeed instead of~\eqref{glkms} let us consider
\be \label{glkms1}
\tilde \Lam_{r} (x) = \Lam_{r} (-x), \qquad \tilde \Phi_{a} (-x) = \Phi_{a} (x) - i \Phi_{r} (x)
\ee
with a minus sign in the transformation of $\Phi_{a}$. With this change all our discussion in Sec.~\ref{sec:theo} goes through except that the explicit form of $S^\mu$ changes into a
new $\hat S^\mu$, which still satisfies
\be
\p_\mu \hat S^\mu = Q \geq 0
\ee
where $Q$ denotes the quantity on the right hand side of~\eqref{fen} and is even under $\Phi_{r} \to - \Phi_{r}$. One finds, however,  that at zeroth derivative order $\hat S^\mu_{0} = - S^\mu_0$. Thus in order to make connection to the standard equilibrium entropy  we
should identify the new entropy current 
as
\be
S_T^\mu = - \hat S^\mu, \quad \Rightarrow \quad \p_\mu S_T^\mu = - Q \leq 0 \ .
\ee
That is, the thermodynamical arrow of time is reversed. One can also further check that dissipative coefficients in
$\sL_{\rm eff}$ are non-negative for~\eqref{glkms} but all switch signs for~\eqref{glkms1}.

The choice of the sign in the $Z_2$ transformation can be traced to a boundary condition on local equilibrium;
the sign in~\eqref{glkms} corresponds to a local equilibrium established in the past, while that in~\eqref{glkms1} to a local equilibrium established in the future. See Appendix~\ref{app:a2} for more discussions.
This is entirely similar to previous derivations of the second law in other contexts: in derivation of the Boltzmann equation and thus Boltzmann's H theorem from the microscopic Liouville equation, the arrow of time depends on whether
 the factorization condition of multiple-particle distribution function is imposed in the past or future~\cite{bogo,cohen,wu}; in various fluctuation theorems,  the second law follows from an initial  equilibrium~(see e.g.~\cite{campisi,Jarz} for reviews).

In addition to the thermodynamic arrow of time, the system also has a causal arrow of time; under a disturbance, the response must come after the cause, not before. In our world the two arrows have always been observed to coincide, as emphasized in~\cite{Evans96}. 
One can readily check that in our setup, for a thermodynamically stable system, the causal arrow and thermodynamical arrow of time do coincide. In particular, under a flip of sign in the $Z_2$ dynamical transformations, the causal arrow of time is also flipped.
See Appendix~\ref{app:a7} for  examples.

The discussion here can be generalized to a number of directions, the most immediate of which is to explore
the consequences of the $Z_2$ $\local$ KMS symmetry inside the path integrals, i.e. to explore the implications of the corresponding ``Ward identities'' at both classical and quantum level.  We expect they should lead to
classical and quantum generalizations of fluctuation relations~\cite{boch,jarz1,crooks,campisi1}.

\vspace{0.2in}   \centerline{\bf{Acknowledgements}} \vspace{0.2in}
We thank P.~Gao, A.~Guth, M.~Kardar, L.~Lussardi, J.~Maldacena, J.~Sonner, M.~Spera, A.~Yarom and J. Zaanen 
for discussion and conversations. Work supported in part by funds provided by the U.S. Department of Energy
(D.O.E.) under cooperative research agreement DE-FG0205ER41360.

\appendix


\section{Constraints from unitarity}  \label{app:a}

Here we provide a derivation of~\eqref{fer1}--\eqref{key1}. Let us consider the path integrals for fast degrees of freedom
\be\label{ori}
A=  \int_{\rho_0} D \psi_1 D \psi_2   \, e^{i S_0 [\psi_1, \chi_1] - i S_0 [ \psi_2, \chi_2]} 
\ee
where $\psi_{1,2}$ are fast variables to be integrated out and $\chi_{1,2}$ are remaining slow variables. We have suppressed external sources which can be trivially added. 
 Let us first consider $\rho_0$ given by a pure state
\be
\rho_0 = \ket{\Psi_0}\bra{\Psi_0}
\ee
where $\Psi_0$ has a wave functional 
$\Psi_0[\psi_1^{(0)},\chi_1^{(0)}]$, where $\psi_1^{(0)}$ and
$\chi_1^{(0)}$ are the initial values of $\chi_1$ and $\psi_1$.  The path integrals~\eqref{ori} can be written more explicitly as
\be \label{ori1}
 \int D \psi_1^{(0)} D \psi_2^{(0)} \int_{\psi_1^{(0)},  \psi_2^{(0)}}^{\psi_1(\infty) = \psi_2 (\infty)} D \psi_1 D \psi_2  \, e^{i S_0 [\psi_1, \chi_1] - i S_0 [ \psi_2, \chi_2]} \, \Psi_0 [\psi_1^{(0)}, \chi_1^{(0)}]
\Psi_0^* [\psi_2^{(0)}, \chi_2^{(0)}] \ .
\ee
Now in the above expression we view all $\chi$'s as external backgrounds for the $\psi$-system, i.e. we can write~\eqref{ori1} as
\be \label{yui}
A =  \vev{\Psi_2 |U^\da_{\rm fast} (\chi_2)  U_{\rm fast} (\chi_1) |\Psi_1}
\ee
where $\ket{\Psi_1}$ has wave functional $\Psi_1 [\psi^{(0)}_1] \equiv \Psi_0 [\psi^{(0)}_1, \chi_1^{(0)}]$ and similarly with $\Psi_2$.
$U_{\rm fast}$ is the evolution operator acting on $\psi$-system with $\chi$ as a background.
Note that
\be
\vev{\Psi_1 |\Psi_1} = f^2 [\chi_1^{(0)}], \qquad \vev{\Psi_2 |\Psi_2} = f^2 [\chi_2^{(0)}]
\ee
where $f$ is some functional of the boundary values of $\chi$ variables.

We  now define $I_{\rm eff}$ as the ``bulk'' part of $A$
\be \label{defi}
A \equiv \rho_0^{\rm eff} [\chi_1^{(0)}, \chi_2^{(0)}]  \, e^{i I_{\rm eff} [\chi_1, \chi_2]}, \qquad  \rho_0^{\rm eff} [\chi_1^{(0)}, \chi_2^{(0)}]
= f [\chi_1^{(0)}] f [\chi_2^{(0)}]
\ee
where $ \rho_0^{\rm eff}$ is interpreted as the effective initial density matrix for slow variables $\chi$.
Now given the unitarity of $U$, we then immediately conclude from~\eqref{yui} and~\eqref{defi} that
\be\label{or2}
|A| \leq \rho_0^{\rm eff} [\chi_1^{(0)}, \chi_2^{(0)}]  \quad \Rightarrow \quad \le|e^{i I_{\rm eff} [\chi_1, \chi_2]} \ri|
 \leq 1 \ .
\ee

The derivation can be readily generalized to a general density matrix $\rho_{0}$ by writing it in a diagonal basis, i.e. $\rho_0 = \sum_n c_n \ket{\Psi_n} \bra{\Psi_n}$ with $\sum_n c_n =1, \; c_n \geq 0$. Then equation~\eqref{yui} becomes
\be\label{yui2}
A = \sum_n c_n  \vev{\Psi_{n2} |U^\da_{\rm fast} (\chi_2)  U_{\rm fast} (\chi_1) |\Psi_{n1}}
\ee
with $\Psi_{n1}, \Psi_{n2}$ defined similarly as before and their normalizations given by $f_n [\chi^{(0)}_{1}], f_n [\chi^{(0)}_{2}]$ respectively. We again use the first equation of~\eqref{defi} to define $I_{\rm eff}$ with now $\rho_0^{\rm eff}$ defined as
\be \label{yui3}
\rho_0^{\rm eff} [\chi_1^{(0)}, \chi_2^{(0)}]
=\sum_n c_n \, f_n [\chi_1^{(0)}] f_n [\chi_2^{(0)}] \ .
\ee
Note that $\rho_0^{\rm eff} $ is properly normalized. From~\eqref{yui} we again have~\eqref{or2} which
 concludes the derivation of ${\rm Im} \;  I_{\rm eff} \geq 0$.

One can further  generalize the above argument by placing the density matrix $\rho_0$ in (\ref{ori}) at finite time $t=t_1$ instead of $t=-\infty$, and close the time path at time $t=t_2$ instead of $t=\infty$, so that the upper boundary condition in the path integrals in (\ref{ori1}) becomes $\psi_1(t_2)=\psi_2(t_2)$.\footnote{In terms of Fig. \ref{fig:SK}(a) this corresponds to take $t_i=t_1$ and $t_f=t_2$.} This leads to the effective action
\be I_{\text{eff}}=\int_{t_1}^{t_2}dt\int d^{d-1}x\,\mathcal L_{\text{eff}}\ ,\ee
for which the above discussion gives
\be\label{posit1} \int_{t_1}^{t_2}dt\int d^{d-1} x \,(\text{Im}\, \mathcal L_{\text{eff}})\geq 0\ .\ee

Taking complex conjugate of~\eqref{defi} we obtain~\eqref{fer1}. Now taking $\chi_1 = \chi_2= \chi$ (including their initial values
$\chi^{(0)}_{1} = \chi^{(0)}_{2} = \chi^{(0)})$, then we find from~\eqref{yui2} and~\eqref{yui3}
\be
A = \sum_n c_n  f_n^2 [\chi^{(0)}] = \rho_0^{\rm eff} [\chi^{(0)}, \chi^{(0)}]
\ee
and thus from~\eqref{defi}
\be
I_{\rm eff} [\chi, \chi] =  0 \
\ee
which gives~\eqref{key1}.

\section{Motivation for dynamical KMS symmetry} \label{app:a2}

\begin{figure}[!h]
\begin{center}
\includegraphics[scale=0.4]{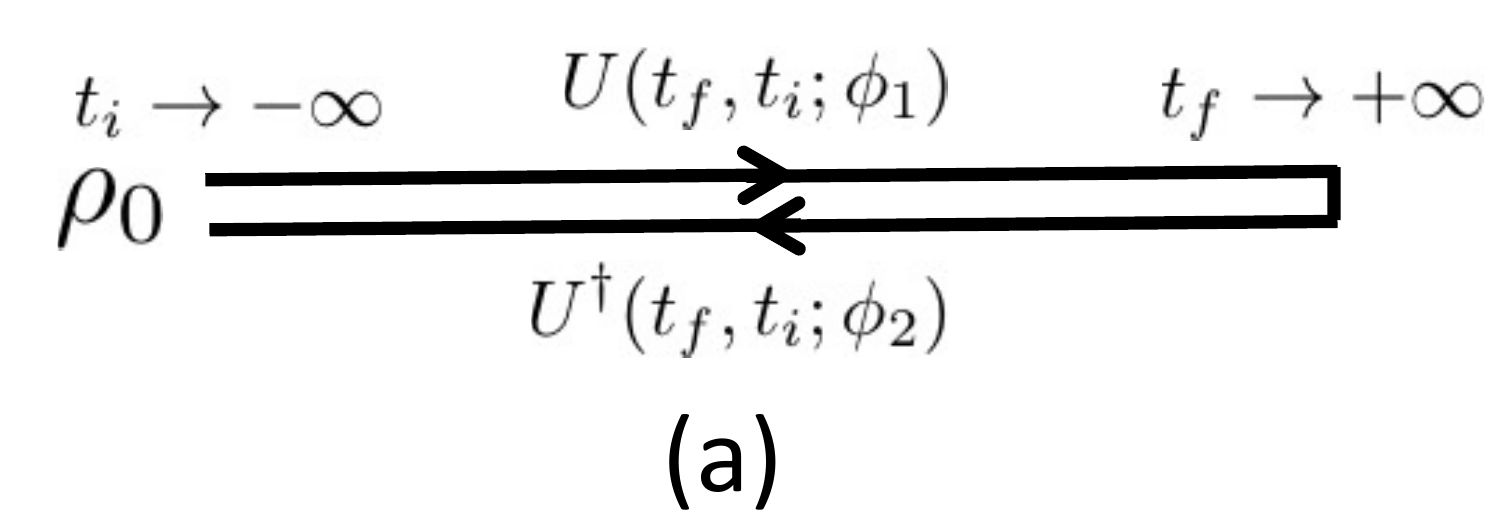} \quad \includegraphics[scale=0.4]{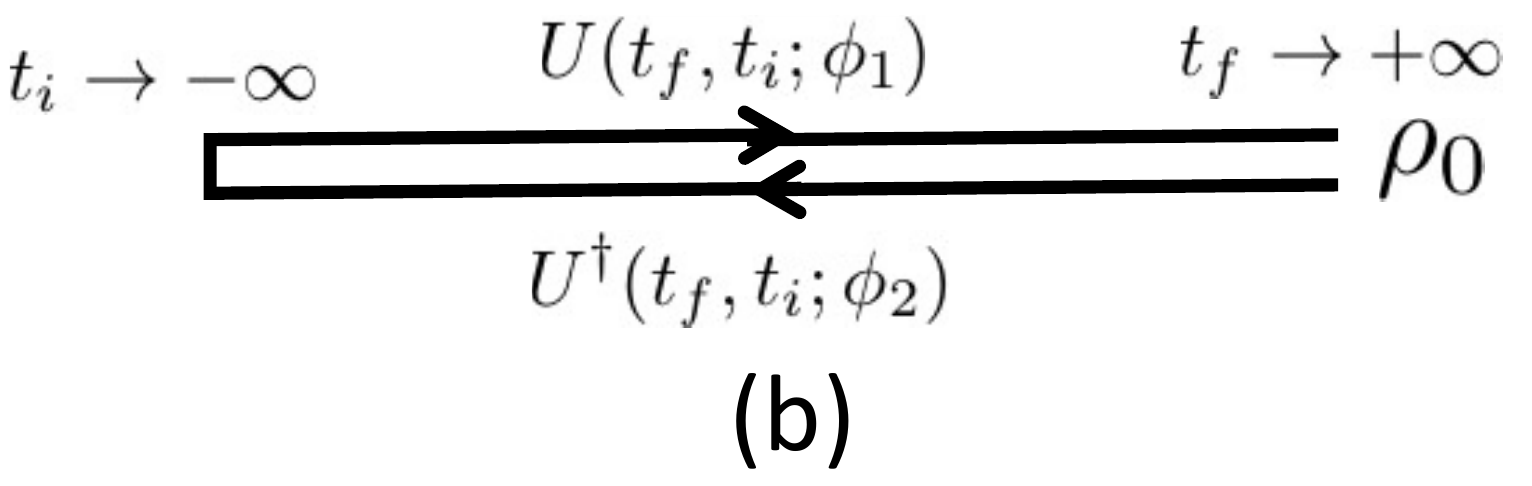}
\end{center}
\caption{ Closed time path contour in a state given by density matrix $\rho_0$. The two lines represent path integrals along two copies of spacetime with arrows indicating the direction of integrations. (a) and (b) correspond to having $\rho_0$ as the initial and final state
respectively.
}
 \label{fig:SK}
\end{figure}

Here we discuss the motivations behind the $Z_2$ symmetry~\eqref{lkms}, and some simple examples.

Now consider CTP generating functional~\eqref{gen0}  with $\rho_0$ given by an initial thermal density matrix with inverse temperature $\beta_0$ (see Fig.~\ref{fig:SK}(a)). Alternatively we can also consider the generating functional $W_T$ defined by Fig.~\ref{fig:SK}(b) with density matrix $\rho_0$ imposed at $t = +\infty$ rather than at $t = -\infty$, i.e.
\bln
e^{W_T [\phi_{1i}, \phi_{2i}]} &=  \Tr\le[U^\da (+\infty, - \infty; \phi_{2i}) \rho_0 U (+\infty, -\infty; \phi_{1i})\ri] \ .
 \label{defwt}
\end{align}
For a system with $\sP \sT$ symmetry then we have\footnote{We again consider real sources. And as discussed in footnote~\ref{foot:2} we choose $\sP \sT$ for definiteness.}
\be \label{tins}
W[\phi_{1i}, \phi_{2i}] = W_T [\phi_{1i}^{PT}, \phi_{2i}^{PT}] ,  \qquad \phi_i^{PT} (x) \equiv \eta_i^{PT} \phi_i (- x)   \
\ee
where $x$ denotes $x^\mu = (x^0, x^i)  = (t, \vx)$.
$W$ and $W_T$ are also related by the Kubo-Martin-Schwinger (KMS) condition~\cite{kubo57,mart59,Kadanoff}
\be \label{tins1}
{W [\phi_{1i}, \phi_{2i}]} = W_T  [\phi_{1i} (t + i \th ), \phi_{2i} (t - i (\beta_0-\th))]
\ee
for $\th \in [0, \beta_0]$ (see Sec. IIC of~\cite{CGL} for more details).
From~\eqref{tins} and~\eqref{tins1} we thus find that
\be
\label{1newfdt1}
W [\phi_1 (x), \phi_2 (x)] = W [\tilde \phi_1 (x); \tilde \phi_2  (x)]
\ee
with
\be
\begin{split}
\tilde \phi_1 (x) = \phi_1 (-t + i \th, - \vx ) , \qquad
\tilde  \phi_2 (x) =  \phi_2 (- t - i (\beta_0 - \th), -\vx )  \
\label{tiV}
\end{split}
\ee
where for simplicity we have taken $\eta^{PT} =1$.
For a theory whose dynamical variables
$\chi$ are non-conserved quantities, the couplings between dynamical variables $\chi$ and the sources $\phi$ can be written
in a linear form
\be
I_{\rm EFT} [\chi_1, \phi_1; \chi_2, \phi_2] = \cdots + \int d^d x \, \le(\chi_1 \phi_1 - \chi_2 \phi_2 \ri) 
 \ .
\ee
It can be readily checked that~\eqref{1newfdt1} is satisfied if we require that $I_{\rm EFT}$ satisfy
\be \label{lkms2}
I_{\rm EFT} [\chi_1, \phi_1; \chi_2, \phi_2] =I_{\rm EFT}   [\tilde \chi_1, \tilde \phi_1; \tilde \chi_2, \tilde \phi_2]
\ee
where $\tilde \phi_{1,2}$ are given by~\eqref{tiV} and
\be
\begin{split}
\tilde \chi_1 (x) = \chi_1 (-t + i \th, - \vx ) , \qquad
\tilde  \chi_2 (x) =  \chi_2 (- t - i (\beta_0 - \th), -\vx )  \ .
\label{tiV1}
\end{split}
\ee
In the classical $ \hbar \to 0$ limit  $\phi_a, \chi_a \to \hbar (\phi_a , \chi_a)$ and $ \phi_r, \chi_r \to \phi_r,  \chi_r$.
Restoring the $\hbar$ in $\beta_0 \hbar, \th \hbar$ in~\eqref{tiV} and~\eqref{tiV1} we then find~\eqref{tiV} and~\eqref{tiV1}
become
\bega \label{2kms}
\tilde \phi_{r} (x) = \phi_r (-x), \qquad \tilde \phi_a (x) = \phi_{a} (-x) + i \beta_0 \p_0 \phi_r (-x) \\
\tilde \chi_{r} (x) = \chi_r (-x), \qquad \tilde \chi_a (x) = \chi_{a} (-x) + i \beta_0 \p_0 \chi_r (-x)\ .
\label{3kms}
\end{gather}
Transformations for $\chi_{r \rmi}, \chi_{a \rmi}$ in~\eqref{glkms} are generalization of~\eqref{3kms} to {\it local} equilibrium.
The transformations~\eqref{glkms} for hydrodynamical variables are discussed in detail in~\cite{CGL1}.

Similarly from~\eqref{tins} and~\eqref{tins1} we find for $W_T$ that
\bega
 W_T [\phi_{1}, \phi_{2}] = W [\phi_1^{PT}, \phi_2^{PT}]  =  W_T  [\phi_{1} (-t - i \th, -\vx ), \phi_{2} (-t + i (\beta_0-\th), -\vx)] \
 \end{gather}
and equations~\eqref{2kms} become
\be \label{bkms1}
\tilde \phi_{r} (x) = \phi_r (-x), \qquad \tilde \phi_a (x) = \phi_{a} (-x) - i \beta_0 \p_0 \phi_r (-x) \\
\ee
i.e. with a minus sign in the second second in the transformation of $\phi_a$. Accordingly the dynamical KMS transformation~\eqref{3kms} should be replaced by
\be \label{glkmsn}
\tilde \chi_{r} (-x) = \chi_r (x), \qquad \tilde \chi_a (-x) = \chi_{a} (x) - i \beta_0 \p_0 \chi_r (x)\ .
\ee

We thus conclude that the sign in transformation of $\Phi_a$ in~\eqref{glkms} may be considered as for a local equilibrium established in the past, the sign in~\eqref{glkms1} may be considered as for a local equilibrium established in the future.



\section{Imposing the dynamical KMS condition} \label{app:a3}

Here we elaborate a bit further on imposing the dynamical KMS conditions~\eqref{gkm1} and~\eqref{gkm2}.
We first note that  there is a simple trick\footnote{Due to Ping Gao, private communication.} to impose conditions~\eqref{gkm2} with $ k \geq 1$,
which also makes manifest that one can set $V_k^\mu$ to zero by absorbing them  into the definition of the Lagrangian.
Consider a Lagrangian density $\sL_0$ of the form~\eqref{w3}. Due to $Z_2$ nature of the transformation,
\be \label{w4}
\sL =\ha \le( \sL_0 + \tilde \sL_0 \ri),
\ee
where $\tilde \sL_0$ is obtained from $\sL_0$ by acting transformations~\eqref{glkms}, automatically satisfies~\eqref{gkm2} without the need for any total derivatives.  Note, however, that $\tilde \sL_0$ in general contains terms with $r$-fields only, and we must then further require that such terms in $\tilde \sL_0$ vanish, which is precisely~\eqref{gkm1}.

For $k=1$ we will need to perform a further integration by parts to write $f^{(1)}$ terms in the Lagrangian in the form of~\eqref{eom11}, with no further derivatives on $a$-variables. This can generate a nonzero $V_1^\mu$. We now show that  $\p_\mu V_1^\mu$ contains only  odd number of derivatives. Acting dynamical KMS transformation on both sides of~\eqref{ryt} we have
\be \label{rty1}
 \mathcal L-\tilde{\mathcal L}=-\p_\mu \tilde V^\mu,
 \ee
where we used that the dynamical KMS transformation is $Z_2$, and that $\tilde V^\mu$ denotes the dynamical KMS
transformed of $V^\mu$. Comparing~\eqref{rty1} with~\eqref{ryt} we find that $\tilde V^\mu = V^\mu$.
Now given that $V^\mu = V_0^\mu + V_1^\mu$, we then conclude that $V_1^\mu$ can only contain even number of derivatives.

\section{Examples} \label{app:a7}

Here we discuss some two explicit examples.

\subsection{Model A}

As an illustration of a system with no conservation laws, we consider the critical dynamics of a $n$-component real order parameter $\chi_\rmi, \rmi=1,\cdots , n$ (i.e. model A~\cite{hohenberg,Folk}). We will ignore couplings to hydrodynamic modes, i.e. the system is at a fixed inverse temperature $\beta_0$ and $\beta^\mu = (\beta_0, \vec 0)$. In~\eqref{w3} $\Lam_r$ and $\Phi_a$
are then $\chi_{r\rmi}$ and $\chi_{a\rmi}$ respectively, and the dynamical KMS transformations~\eqref{glkms} become
\bega \label{ckms1}
\tilde \chi_{r \rmi} (x)  = \chi_{r  \rmi}(-x) , \qquad \tilde \chi_{a \rmi} (- x) = \chi_{a \rmi} (x) +{ i \beta_0  } \p_0 \chi_{r \rmi} (x) \ . 
\end{gather}
As~\eqref{ckms1} only involves time derivative we can treat time and spatial derivatives separately. For simplicity we consider the first two terms in~\eqref{w3} which can be written explicitly as
\be \label{sl1}
\sL = E^\rmi \chi_{a \rmi} +i   X^{\rmi \rmj} \chi_{a \rmi} \chi_{a \rmj} 
+ \cdots  \
\ee
where one should keep in mind that $X^{\rmi \rmj}$ may include derivatives on $\chi_{a}$'s and thus do not have to be symmetric
in exchanging $\rmi, \rmj$ indices. We can expand $E$ and $X$ in the number of time derivatives as
\be
E^\rmi = E_0^\rmi + E^\rmi_1 + \cdots , \qquad X^{\rmi \rmj} =X^{\rmi \rmj}_0 + X^{\rmi \rmj}_1 + \cdots
\ee
and each term can be further expanded in terms of the number of spatial derivatives.

Applying~\eqref{w4} to~\eqref{sl1} we can read the consequences of~\eqref{gkm2}
\be \label{po1}
 E_1^\rmi = - \ha ( X^{\rmi \rmj }_0 +  X^{\rmj \rmi }_0) \Phi_{\rmj} , \quad
\Phi_\rmi \equiv \beta_0 \p_0 \chi_{r \rmi} \ .
\ee
Now for simplicity let us further restrict to zero spatial derivative in $X_0$, i.e.
\be \label{ol}
X_{0}^{\rmi \rmj}  = f^{\rmi \rmj} (\chi_r)    \quad \Rightarrow \quad E_{1}^\rmi = - f^{\rmi \rmj} (\chi_r) \Phi_{\rmj}  \ .
\ee
$f^{\rmi \rmj}$ is symmetric in its indices, which leads to the Onsager relations.
Applying~\eqref{gkm1} to~\eqref{sl1}, we find
\bega \label{f1}
 E^\rmi_0 \Phi_\rmi  = \p_\mu V^\mu_{(0,0)} ,  \qquad
 - E^{\rmi}_1 \Phi_\rmi - X^{\rmi \rmj}_0 \Phi_\rmi \Phi_\rmj  =  \p_\mu V^\mu_{(0,1)}
 \end{gather}
where $V^\mu_{(0,n)}$ contains $n$ time derivatives. The second equation is automatically satisfied with $V^\mu_{(0,1)} =0$ due to~\eqref{ol}. The first equation can be solved to all orders in spatial derivatives  if there exists a local functional
$\sF (t; \chi_r]$ (in~\eqref{ii1} there are only spatial integrations)
\be \label{ii1}
\sF (t; \chi_r] = \int d^{d-1} \vec x \, F (\chi_r (x), \p_i \chi_r (x), \cdots)
\ee
from which
\be \label{ii2}
E_0^\rmi
= - {\de \sF \ov \de \chi_{r \rmi} (x)}  = - { \p F \ov \p \chi_{r \rmi}} + \p_i \le({\p F \ov \p \p_i \chi_{r \rmi}}\ri) -
 \p_i^2 \le({\p F \ov \p \p_i^2 \chi_{r \rmi}} \ri)+ \cdots
 \ee
and accordingly
\be \label{vmu}
V^0_{(0,0)} = - \beta_0 F , \qquad {1 \ov \beta_0} V^i_{(0,0)} = {\p F \ov \p \p_i \chi_{r \rmi}} \p_0 \chi_{r \rmi} + {\p F \ov \p \p_i^2 \chi_{r \rmi}} \p_i \p_0 \chi_{r \rmi} -  \p_i {\p F \ov \p \p_i^2 \chi_{r \rmi}} \p_0 \chi_{r \rmi}  + \cdots \ .
\ee
Collecting various expressions above, we can write the Lagrangian as
\be \label{lag1}
\sL_{\rm eff}=\le(- {\de \sF \ov \de \chi_{r \rmi}}  -  \beta_0 f^{\rmi \rmj}  \p_0 \chi_{r \rmj} \ri) \chi_{a \rmi}
+ {i } f^{\rmi \rmj} (\chi_r)   \chi_{a \rmi}   \chi_{a \rmj }  +  \cdots \ .
\ee
Equation~\eqref{pos} also requires that for arbitrary $a_\rmi (x) $
\be \label{posi1}
f^{\rmi \rmj} (\chi_r)  a_\rmi (x) a_\rmj (x) \geq 0 \ .
\ee

We can now readily write the entropy current to the order exhibited in~\eqref{lag1} by applying equation~\eqref{cse}, which gives
\be
S^\mu = V^\mu_0 \ .
\ee
More explicitly,
\be\label{ss0}
S^0  = - \beta_0 F, \qquad S^i =  \beta_0 \le( {\p F \ov \p \p_i \chi_{r \rmi}} \p_0 \chi_{r \rmi} + {\p F \ov \p \p_i^2 \chi_{r \rmi}} \p_i \p_0 \chi_{r \rmi} -  \p_i {\p F \ov \p \p_i^2 \chi_{r \rmi}} \p_0 \chi_{r \rmi}  + \cdots \ri)
\ee
and one can readily check after using equations of motion
 \be
  \p_\mu S^\mu =  \beta_0^2 f_{\rmi \rmj}  \p_0 \chi_{r \rmj} \p_0 \chi_{r \rmi} \geq 0 \  .
 \ee
At zeroth order in time derivatives we have
\be
S_0^0 = - \beta_0 F, \qquad S^i_0 = 0 \
\ee
which has the standard form with $F$ interpreted as the (static) free energy density of the scalar system.

Let us now consider a phase whose equilibrium configuration has $\chi_{r \rmi} =0$ and $\chi_{r\rmi}, \chi_{a \rmi}$ are small.   Keeping only quadratic terms in~\eqref{lag1}, we can write
\be
f^{\rmi \rmj} = {1 \ov \Ga_0 } \de^{\rmi \rmj}  , \qquad
F =  \ha r \chi_{r \rmi}^2 + \ha  (\p_i \chi_{\rmi})^2 + \cdots
\ee
where we have only kept two spatial derivatives in $F$.  $\Ga_0$ should be non-negative due to~\eqref{posi1}.
$F$ (and thus the constant $r$) should also be non-negative to ensure thermodynamic stability.\footnote{We emphasize that this non-negativity, which concerns whether the equilibrium state itself is a stable phase,
has nothing to do with~\eqref{pos} which concerns with dynamics.} With an external source $\phi_{r}$, the quadratic Lagrangian can be written as
\be
\sL = \le( - r \chi_{r \rmi} +  \p_i^2 \chi_{r \rmi}  -  {\beta_0 \ov \Ga_0}  \p_0 \chi_{r \rmj}  \ri) \chi_{a \rmi}
+ {i \ov \Ga_0 }  \chi_{a \rmi}   \chi_{a \rmi}  + \phi_{r\rmi} \chi_{a \rmi}  \
\ee
and the equations of motion for $\chi_{r \rmi}$ are
 \be
\ga_0  \p_0 \chi_{r \rmi} +  r  \chi_{r \rmi}  -  \p_i^2 \chi_{r \rmi} = \phi_{r \rmi}  \
\ee
where the ``friction'' coefficient
\be \label{0}
\ga_0 = \beta_0 \Ga_0 \
\ee
is also non-negative, i.e.  $\chi_{r \rmi}$ will be damped. Equivalently we find the response function in momentum space
\be \label{cou}
\chi_{r \rmi} (\om, \vk)  = {\phi_{r \rmi} (\om, \vk) \ov - i \ga_0 \om + r + k^2 }
\ee
has a pole in the lower half complex $\om$-plane.

Now consider changing the sign of the second term in~\eqref{ckms1}, which may be considered as taking $\beta_0 \to - \beta_0$ we then find that  the new current which we denotes as $\hat S^\mu$ has opposite signs to~\eqref{ss0}, but still satisfies
\be
\p_\mu \hat S^\mu =  \beta_0^2 f_{\rmi \rmj}  \p_0 \chi_{r \rmj} \p_0 \chi_{r \rmi} \geq 0 \  .
\ee
To match with the standard equilibrium expression for the entropy density, we then need to identify the new entropy current $S^\mu_T$ as $- \hat S^\mu$, which then satisfies
\be
\p_\mu S^\mu_T = - \beta_0^2 f_{\rmi \rmj}  \p_0 \chi_{r \rmj} \p_0 \chi_{r \rmi} \leq 0 \  .
\ee
In~\eqref{0}, $\ga_0$ also changes sign and now the pole of~\eqref{cou} lies
in the upper half plane. Thus both thermodynamic and causal arrows of time switch.

\subsection{Fluctuating hydrodynamics for relativistic charged fluids}

We now briefly outline the story for fluctuating hydrodynamics of a relativistic charged fluids in the classical limit.
More details can be found~\cite{CGL1}.

The dynamical variables are then hydrodynamical modes associated with conserved quantities. The $r$-variables
$\beta^\mu = \beta (x) u^\mu (x)$ and $\hmu = \beta (x) \mu (x)$ can be written in a uniform manner as
$\beta_M = (\beta_\mu, \hat \mu)$. 
The $a$-variables can be written in a uniform manner as $ X_{a M} = (X_{a \mu}, \vp_a)$.
Equation~\eqref{glkms} can also be written uniformly as
\be \label{oun}
 \p_\mu \tilde X_{a M} (-x) = \p_\mu \tilde X_{a M} (x)
+i \p_\mu \beta_M (x)
\ee
and the first two terms of~\eqref{w3}  can be written more explicitly as
\be \label{inp}
\sL_{\rm eff} = T^{\mu M} \p_\mu X_{a M} + i W^{\mu  \nu, M N} \p_\mu X_{a M} \p_\nu X_{a N}
 + \cdots \ .
\ee
with $T^{\mu M} = (T^{\mu \nu} , J^\nu)$ the hydrodynamic stress tensor and $U(1)$ current. 
We will consider~\eqref{inp} to one derivatives in $T^{\mu M}$ and zero derivative in $W^{\mu \nu, MN}$.

 Applying~\eqref{w4} we find
\bega \label{uio0}
T^{\mu M}_1  = -  W^{\mu \nu, MN}_0 \p_\nu \beta_N
\end{gather}
and equation~\eqref{gkm1} requires
\be \label{003}
T^{\mu M}_0 \p_\mu \beta_M = \p_\mu V^\mu_{(0,0)} , \quad
T^{\mu M }_1 \p_\mu \beta_M +  W^{\mu \nu, MN }_0 \p_\mu \beta_M \p_\nu \beta_N
= \p_\mu V^\mu_{(0,1)}
\ee
where subscripts in $T$ and $W$ now denote the total number of derivatives and so does the second subscript of $V^\mu$.
Note from~\eqref{uio0} the second equation of~\eqref{003} is automatically satisfied  with $V^\mu_{(0,1)} =0$.

At zeroth derivative order, we can write $T^{\mu M}_0  = (T_0^{\mu \nu}, J^\mu_0)$ as
\be
T_0^{\mu \nu}= \varepsilon_0 u^\mu u^\nu+p_0\Delta^{\mu\nu},\quad J^\mu_0 =n_0 u^\mu   \ 
\ee
where $\ep_0, p_0, n_0$ are functions of $\beta$ and $\hmu$.
The first equation of~\eqref{003} then requires $\ep_0, p_0, n_0$ satisfy the standard thermodynamic relations
\be \label{thermo}
\ep_0 + p_0  = -\beta {\p p_0 \ov \p \beta} , \qquad
n_0 = \beta {\p p_0 \ov \p \hmu}   ,
\ee
with
\be\label{entr1}
V_{(0,0)}^\mu= p_0 \beta^\mu \ .
\ee
In other words, the first law of thermodynamics is satisfied locally.
Equation~\eqref{uio0} ensures that $T^{\mu M}$ satisfies the Onsager relations due to
$W^{\mu \nu, MN}_0= W^{ \nu \mu, NM}_0$.

From~\eqref{cse} the entropy current to first derivative order can be written as
\be \label{zeen}
S^\mu
=  p_0 \beta^\mu  -  T^{\mu \nu} \beta_\mu -  J^\mu \hmu \
\ee
and one can readily check that by using equations of motion
\be
\p_\mu S^\mu = W_0^{\mu\nu,MN}\nabla_\mu\beta_M\nabla_\nu \beta_N \geq 0  \ .
\ee
With a bit more effort the right hand side of the above equation can be expressed in a conventional form using
conductivity, shear viscosity and bulk viscosity, see~\cite{CGL1}, where we also generalize the above entropy current
analysis to second order in derivative expansion.

Linear responses from the effective action~\eqref{inp} have
been discussed in details in~\cite{CGL}. Here we only mention some key elements. The response functions have  poles only in the
lower half $\om$-plane\footnote{We again assume the equilibrium phase is thermodynamically stable.}  provided that the leading dissipative coefficients, which are conductivity $\sig$, shear viscosity $\eta$, and  bulk viscosity $\ka$, are all non-negative.
These dissipative coefficients are indeed non-negative as they can be expressed via~\eqref{uio0}
schematically as
\be \label{trd}
\sigma = \beta A_1, \qquad \eta = \beta A_2 , \qquad \ka = \beta A_3
\ee
where $A_{1,2,3}$  are combinations coefficients of $W_0$ and are non-negative separately from~\eqref{pos}.

Now let us consider reverse the sign in~\eqref{oun}, which can be achieved by taking $\beta \to - \beta$ in various places. The resulting $\hat S^\mu_0$ has an opposite overall sign to~\eqref{zeen} and satisfy $\p_\mu \hat S^\mu \geq 0$.
Matching with the standard thermodynamic entropy we should identify $S_T^\mu = - \hat S^\mu$, which then has a negative divergence. Similarly all the dissipative coefficients in~\eqref{trd} change signs and now the poles of response functions lie in upper half frequency plane.

\section{Details of proof} \label{app:a4}

In this Appendix we provide details for the manipulations from~\eqref{c1} to~\eqref{c4}.
First we verify that~\eqref{Ff1}--\eqref{Ff2} solve~\eqref{c1}--\eqref{c2}.
Plugging~\eqref{Ff1} into~\eqref{c1} and rearranging the double sum of the second term on left hand side we find that
\bega
{1 \ov 2m+1} \sum_{k=0}^{l}  (-1)^{k}   C_{2m+2k+2}^{2m} G_{2 k+2}  F^{(2m+2k+2, 2l-2k)}  \cr
  \ha {1 \ov 2m+1} \sum_{k=1}^{l}  (-1)^{k} C_{2m+2k+2}^{2m} F^{(2m+2k+2,2l-2k)}  \le( \sum_{n=1}^{k} C_{2k+2}^{2n} G_{2n} \ri)
\cr
= \ha  \sum_{k=0}^l (-1)^{k+1} C_{2m+2k+2}^{2k+1} F^{(2m+2k+2,2l-2k)}  \ .
\end{gather}
$k=0$ term in the above equation is satisfied as $G_2 =-1$. The terms with $1 \leq k \leq l$ are satisfied from the identities (equation~\eqref{gen1} of Appendix~\ref{app:Ber}) 
\be
  -  G_{2 k+2}     - {1 \ov 2} \le( \sum_{n=1}^{k} C_{2k+2}^{2n} G_{2n} \ri) = k  + 1, \qquad k=1,2, \cdots \ .
  \ee
Similarly plugging~\eqref{Ff2} into~\eqref{c2} and rearranging the double sum on the left hand side we find
\bega
{2 \ov 2m-1} \sum_{k=0}^l  (-1)^k C_{2m + 2k}^{2m-2} F^{(2m+2k, 2l-2k+1)} \le(\sum_{n=0}^{k} C_{2k+2}^{2n} B_{2n} \ri)
=  \sum_{k=0}^{l} (-1)^{k} C_{2m+2k}^{2m-1} F^{(2m+2k,2l-2k+1)}
\end{gather}
which are indeed satisfied given the identities (equation~\eqref{Ber2} of Appendix~\ref{app:Ber})
\be
{1 \ov k +1} \sum_{n=0}^{k} C_{2k+2}^{2n} B_{2n}  = 1, \qquad k =0,1,\cdots\ .
\ee


Now let us give intermediate steps leading to~\eqref{c4}. Plugging~\eqref{Ff1}--\eqref{Ff2} into~\eqref{cse1}, we
find for $n \geq 1$
\bega
  \p_\mu  S_{2n-1}^\mu  = F^{(2, 2n-2)}  + \sum_{k=2}^n o_k F^{(2k, 2n-2k)} , \qquad
 \p_\mu  S_{2n}^\mu = \sum_{k=1}^n e_k F^{(2k, 2n-2k+1)}
\end{gather}
 with
 \be \label{eo0}
 o_k =(-1)^{k+1} \le(1 +{1 \ov 2k+1} \sum_{l=1}^{k-1}  C_{2k+1}^{2l} G_{2l} \ri) , \quad e_k =
 (-1)^{k+1} \le(1 - {2 \ov 2k+1} \sum_{l=0}^{k-1}  C_{2k+1}^{2l} B_{2l} \ri)
   \ .
 \ee
 Using the identities~\eqref{gen2} and~\eqref{Ber1} of Appendix~\ref{app:Ber}
 we find
 \be
 o_k = (-1)^k G_{2k}, \qquad e_k = 2 (-1)^{k+1} B_{2k}
 \ee
 which then give~\eqref{c4}.


To conclude this subsection let us elucidate the structure of equations~\eqref{c1}--\eqref{c2}
which can be rewritten as two infinite families of upper triangular linear equations.
More explicitly, for each integer $l \geq 1$, introducing column vectors
\be
A^{(l)} = \bma F^{(3, 2l-1)} \cr \vdots \cr F^{(2l+1,1)} \ema, \quad
B^{(l)}=  \bma F^{(4, 2l-2)} \cr \vdots \cr F^{(2l+2,0)} \ema, \quad
C^{(l)} = \bma F^{(3, 2l-2)} \cr \vdots \cr F^{(2l+1,0)} \ema, \quad
D^{(l)}=  \bma F^{(2, 2l-1)} \cr \vdots \cr F^{(2l,1)} \ema,
\ee
i.e. for $m = 1, \cdots , l$
\be
A_m^{(l)} = F^{(2m+1, 2l-2m+1)} , \quad B_m^{(l)} = F^{(2m+2, 2l-2m)} , \quad
C_m^{(l)} = F^{(2m+1, 2l-2m)} , \quad D_m^{(l)} = F^{(2m, 2l-2m+1)},
\ee
then we can write~\eqref{c1}--\eqref{c2} as
\be
K^{(l)} \cdot A^{(l)} = L^{(l)} \cdot  B^{(l)}, \qquad M^{(l)} \cdot C^{(l)} = N^{(l)} \cdot  D^{(l)}
\ee
where $K^{(l)}, L^{(l)},M^{(l)},N^{(l)}$ are $l \times l$ upper triangular matrices. Their non-vanishing matrix elements are given by
\bega
 K^{(l)}_{m,m+k}= \de_{k,0} + \ha (-1)^k C_{2m+1+2k}^{2m+1} , \quad
 L^{(l)}_{m,m+k}= \ha (-1)^{k+1} C_{2m+2k+2}^{2m+1} ,  \\
 M^{(l)}_{m,m+k}= (-1)^{k+1} C_{2m+2k+1}^{2m-1}, \quad N^{(l)}_{m,m+k}= (-1)^{k+1} C_{2m+2k}^{2m-1}
\end{gather}
where  $m =1,2 , \cdots, l$ and $k=0,1 , \cdots l-m$. The solution~\eqref{Ff1}--\eqref{Ff2} implies the identities\footnote{We have not found the appearance of these identities in the literature.}
\be
(K^{(l)})^{-1} L^{(l)} = P^{(l)}, \qquad  (M^{(l)})^{-1} N^{(l)} = Q^{(l)}
\ee
where $P^{(l)}, Q^{(l)}$ are upper triangular matrices with nonzero entries given by
\be
P^{(l)}_{m,m+k}= {(-1)^{k} \ov 2m+1}  C_{2m+2k+2}^{2m} G_{2k+2}, \qquad
Q^{(l)}_{m,m+k}= {2 (-1)^{k} \ov 2m+1}  C_{2m+2k}^{2k} B_{2k}
\ee
for $m =1,2 , \cdots, l$ and $k=0,1 , \cdots l-m$.

\subsection{Bernoulli and Genocchi numbers} \label{app:Ber}

Here we collect some facts and identities regarding Bernoulli and Genocchi numbers.
Firstly note the following recursion relations  for Bernouli and numbers~\cite{bernouli,genocchi}
\bega \label{bb1}
B_m = 1 - \sum_{k=0}^{m-1} C_m^k {B_k \ov m- k +1}  = 1 - {1 \ov m+1} \sum_{k=0}^{m-1} C_{m+1}^k B_k,  \quad m \geq 1
\quad {\rm with} \; B_1 = \ha , \\
2 G_m + \sum_{k=1}^{m-1} C_m^k G_k =0, \quad m \geq 2, \quad {\rm with} \;\; G_1 = 1 \ .
\label{bb2}
\end{gather}
Taking $m=2k$ and $m=2k+1$ respectively in~\eqref{bb1} we have
\bega \label{Ber1}
B_{2k} = \ha - {1 \ov 2k +1} \sum_{l=0}^{k-1} C_{2k +1}^{2l} B_{2l},  \quad k =1,2, \cdots \\
\label{Ber2}
0 = \ha - {1 \ov 2 k} \sum_{l=0}^{k-1} C_{2k}^{2l} B_{2l} ,  \quad k =1,2, \cdots  \ .
\end{gather}
Similarly taking $m=2k$ and $m=2k+1$ respectively in~\eqref{bb2} we find
\bega
 \label{gen1}
G_{2k} + k + \ha \sum_{l=1}^{k-1} C_{2k}^{2l} G_{2l} = 0 , \quad k =1,2, \cdots \\
\label{gen2}
\sum_{l=1}^{2k} C_{2k+1}^l G_l =0 \quad \to \quad G_{2k} + 1 + {1 \ov 2k+1}\sum_{l=1}^{k-1} C_{2k+1}^{2l} G_{2l} = 0 ,
\quad k =1,2, \cdots \ .
\end{gather}

\subsection{Two identities} \label{app:a6}

Consider the function
\be\label{fx1} f(x)=\sum_{k=1}^\infty c_{2k} x^{2k},\ee
such that $f(x)$ goes to zero sufficiently fast as $x\to \pm\infty$. Then
\bega
\label{g11} \sum_{k=1}^\infty (-1)^k G_{2k} c_{2k}=\pi \int_{-\infty}^\infty \frac{\cosh (\pi x)}{\sinh^2 (\pi x)}f(x)dx,\\
\label{g12} 2 \sum_{k=1}^\infty (-1)^{k+1} B_{2k} c_{2k}=\pi  \int_{-\infty}^\infty \frac{1}{\sinh^2 (\pi x)}f(x)dx
\end{gather}
which follow from the integral representations of the Bernoulli numbers~\cite{wolfram}
\be
2  (-1)^{k+1} B_{2k} = \pi  \int_{-\infty}^\infty dx \, \frac{1}{\sinh^2 (\pi x)} x^{2k}, \quad
(-1)^k G_{2k} =\pi \int_{-\infty}^\infty d x \, \frac{\cosh (\pi x)}{\sinh^2 (\pi x)} x^{2k}  \ .
\ee

\section{Non-negativity of zeroth order Lagrangian} \label{app:lag}

Consider an action
\be 
S = \int d^d x \, \sL \le(\phi (x), \p_\mu \phi (x), \cdots \ri)  \ .
\ee
Now suppose we have 
\be 
S \geq 0
\ee
for any choice of $\phi (x)$. We would like to show that 
\be 
\sL_0 \geq 0
\ee
for any $\phi$, where $\sL_0$ denote the zero derivative part of $\sL$. 

Take $\phi (x)= \phi_0$ where $\phi_0$ is any constant, then we have 
\be 
S = V \sL_0 (\phi_0) \geq 0 \quad \to \quad \sL_0 (\phi_0) \geq 0
\ee
where $V$ here denotes the spacetime volume. We thus find 
\be 
\sL_0 (\phi (x)) \geq 0
\ee
for any $x$. 

One can also readily see from $S \geq 0$ one cannot conclude the full Lagrangian density to be non-negative. 
Consider adding to $\sL$ a total derivative 
\be 
\sL' = \sL + \p_\mu V^\mu \ . 
\ee
$S$ does not change, but at  a given point $x$ it appears that no matter what the value of $\sL$ is we can always arrange $V^\mu$ to make $\sL'$ to be negative. Note that adding a total derivative does not change $\sL_0$. 

One could contemplate  whether it is possible to use the freedom of adding total derivatives to choose to a Lagrangian $\sL \geq 0$. It is not clear to us whether this is possible or not.


\end{document}